\documentclass[aps,twocolumn,showpacs,preprintnumbers,amsmath,amssymb,nofootinbib,superscriptaddress,showkeys]{revtex4}

\usepackage{graphicx}

\def\slashchar#1{\setbox0=\hbox{$#1$}
   \dimen0=\wd0 \setbox1=\hbox{/} \dimen1=\wd1
   \ifdim\dimen0>\dimen1 \rlap{\hbox to \dimen0{\hfil/\hfil}} #1
   \else  \rlap{\hbox to \dimen1{\hfil$#1$\hfil}} / \fi}

\begin{document}

\title{Pion transition form factor in the Regge approach and
  incomplete vector-meson dominance~\footnote{Supported by Polish
    Ministry of Science and Higher Education grant N202~249235,
    Spanish DGI and FEDER funds with grant FIS2008-01143/FIS, Junta de
    Andaluc{\'\i}a grant FQM225-05, and EU Integrated Infrastructure
    Initiative Hadron Physics Project contract RII3-CT-2004-506078.}
}
\author{Enrique Ruiz Arriola}\email{earriola@ugr.es}
\affiliation{Departamento de F\'{\i}sica At\'omica, Molecular y
  Nuclear, Universidad de Granada, E-18071 Granada, Spain}
\author{Wojciech Broniowski} \email{Wojciech.Broniowski@ifj.edu.pl}
\affiliation{The H. Niewodnicza\'nski Institute of Nuclear Physics,
  PL-31342~Krak\'ow, Poland} \affiliation{Institute of Physics, Jan
  Kochanowski University, PL-25406~Kielce, Poland}

\date{5 April 2010}

\begin{abstract} 
 The concept of incomplete vector-meson dominance and Regge models is
 applied to the transition form factor of the pion.  First, we argue
 that variants of the chiral quark model fulfilling the chiral
 anomaly may violate the Terazawa-West unitarity bounds, as these
 bounds are based on unverified assumptions for the real parts of the
 amplitudes, precluding a possible presence of polynomial terms.  A
 direct consequence is that the transition form factor need not
 necessarily vanish at large values of the photon
 virtuality. Moreover, in the range of the BaBar experiment, the
 Terazawa-West bound is an order of magnitude above the data, thus is
 of formal rather than practical interest. Then we demonstrate how the
 experimental data may be properly explained with incomplete
 vector-meson dominance in a simple model with one state, as well as
 in more sophisticated Regge models. Generalizations of the simple
 Regge model along the lines of Dominguez result in a proper
 description of the data, where one may adjust the parameters in such
 a way that the Terazawa-West bound is satisfied or violated.  We
 also impose the experimental constraint from the $Z \to \pi_0
 \gamma$ decay. Finally, we point out that the photon momentum asymmetry parameter may
 noticeably influence the precision analysis.
\end{abstract}



\maketitle

\section{Introduction \label{sec:intro}}

The pion transition form factor, $F_{\pi^0 \gamma \gamma^\ast}$,
measured in the $e^+ e^- \to \pi^0 \gamma$ annihilation, has been a
particularly interesting object of study since its value at the origin
is fixed by the chiral anomaly~\cite{Adler:1969gk,Bell:1969ts} while
the behavior at very high Euclidean momenta has been predicted via
perturbative quantum chromodynamics (pQCD), apparently holding at
sufficiently high but unspecified
scales~\cite{Lepage:1979zb,Efremov:1979qk}. However, objections have
been raised~\cite{Isgur:1984jm} and reiterated \cite{Isgur:1988iw}
against the applicability of pQCD to exclusive processes.  The recent
measurement of $F_{\pi^0 \gamma \gamma^\ast}$ by the BaBar
Collaboration~\cite{Aubert:2009mc}, where the pion transition form
factor goes visibly above the conventional pQCD prediction at scales
$Q^2 > 15~{\rm GeV}^2$, raised serious doubts concerning the
theoretical understanding of the exclusive processes, also in the
asymptotic region. In the pQCD approach based on factorization, one
uses the light-cone Feynman rules and the ERBL evolution of the pion
distribution amplitude (PDA)
\cite{Efremov:1979qk,Lepage:1979zb,Lepage:1980fj,Brodsky:1981rp,delAguila:1981nk,Braaten:1982yp,Kadantseva:1985kb,%
  Bakulev:2001pa,Bakulev:2002uc,Bakulev:2003cs,Mikhailov:2009kf},
which results asymptotically in the leading-twist Brodsky-Lepage term
$F_{\pi^0 \gamma \gamma^\ast} \sim 2 f_\pi/Q^2$, in vivid
contradiction to the recent BaBar~\cite{Aubert:2009mc} data.

Several ideas have been proposed to circumvent the
problem. Radyushkin~\cite{Radyushkin:2009zg} pointed out that the
presence of the possible end-point singularities in the PDA (as found
by the authors in the chiral quark models at the low-energy
quark-model scale \cite{RuizArriola:2002bp}), together with
essentially switched-off evolution and regulated quark propagators, is
capable of reproducing the data in the available $Q^2$ domain. Similar
conclusions were reached by Polyakov~\cite{Polyakov:2009je}. In this
approach asymptotically $F_{\pi^0 \gamma \gamma^\ast} \sim
\log(Q^2/\mu^2)/Q^2$, with the $\log$ in the numerator indicating the
breaking of factorization. We note that the same asymptotics follows
in the Spectral Quark Model (SQM), see Eq.~(14.1) of
Ref.~\cite{RuizArriola:2003bs}.  In fact, serious concerns have been
spoken out on the validity of factorization~\cite{Mikhailov:2009sa} in
the considered process.
Dorokhov~\cite{Dorokhov:2009dg,Dorokhov:2009zx,Dorokhov:2009jd}
proposed the use of the fixed-mass constituent quark model to evaluate
the triangle diagram of Fig.~\ref{fig:tri}, which is capable of reproducing the BaBar data,
however the needed value of the constituent quark mass is
uncomfortably low, $M \sim 135$~MeV. In this model the asymptotics has
the form $\sim [\log(Q^2/\mu^2)]^2/Q^2$.  Possible need of the
higher-twist terms has been invoked in Ref.~\cite{Noguera:2010fe}.
The calculation of Kotko and Praszalowicz~\cite{Kotko:2009ij} in the
non-local chiral quark model inspired by the instanton-liquid model of
QCD produced the result in agreement with the data at lower values of
$Q^2$. The influence of non-perturbative gluonic components of the
pion stemming from instantons has been considered in
Ref.~\cite{Kochelev:2009nz}. Mikhailov and Stefanis~\cite{Mikhailov:2009kf} showed that the 
significant experimental 
growth of the transition form factor between 10 and 40 GeV$^2$ cannot be explained in terms of
higher-order pQCD corrections at the NNLO level.
Finally, in a very recent
paper~\cite{Dorokhov:2010bz} Dorokhov reiterates the findings of
Ref.~\cite{Dorokhov:2009dg} in the non-local model, where the
asymptotic behavior is $\sim \log(Q^2/\mu^2)/Q^2$. The non-local model
of Ref.~\cite{Dorokhov:2010bz} also requires a very low constituent
quark mass in order to reproduce the data in the whole $Q^2$ range.

Given the fact that the recent BaBar data~\cite{Aubert:2009mc} predate
the conventional pQCD expectations at such high virtualities as
$15-30~{\rm GeV}^2$, it seems adequate to question any implicit
assumptions in theoretical analyses.  In this paper we analyze
critically some of the assumptions underlying certain high-energy
theorems which, if holding, would forbid the simple pre-BaBar fits based on
the incomplete vector-meson dominance (IVMD)~\cite{Knecht:2001xc} (see
also Ref.~\cite{Nyffeler:2009uw} where this result is upgraded with
tiny consequences for to the muon $g-2$). The theorems in question, 
the Terazawa-West (TW) bounds~\cite{Terazawa:1973hk,West:1973gd}, 
were recalled by Dorokhov in Ref.~\cite{Dorokhov:2009jd}.

The TW bounds~\cite{Terazawa:1973hk,West:1973gd} (for a review see,
e.g.,~\cite{Terazawa:1973tb}) are central to our 
discussion; they are reviewed in Sec.~\ref{sec:TW}.  
Their derivation uses the Schwarz
inequality to predict that for physical momenta ${\rm Im} F_{\pi^0
  \gamma \gamma^\ast} = {\cal O}(1/\sqrt{q^2})$ (the first bound). If
one further {\em assumes} that there are no polynomial contributions
to the real part of the form factor, one may conclude with the help of
a dispersion relation that $F_{\pi^0 \gamma \gamma^\ast}= {\cal
  O}(1/Q)$ for space-like momenta $Q$ (the second bound). The
assumption concerning the real part is clearly spelled out in the
original works and, to our knowledge, has never been credibly
questioned.  We should remind the reader that this was a pre-QCD
bound, so we may test the validity of the assumption by analyzing a
particular model where {\it both} the anomaly is satisfied and the
second Terazawa-West bound is violated.  Actually, a reanalysis of the
issue using the quark model and corresponding triangle graphs was
presented in Ref.~\cite{West:1990ta} as an example case where the
expected bound behavior holds. Here we will consider another variant
of the quark model which in fact does not fulfill the assumptions,
hence leads to a violation of the second bound. The key point is to
realize that the anomaly fixes the value of the form factor at the
origin, while the finite $Q$ behavior can be determined regardless of
the value at $Q=0$. This construction requires a subtraction constant
which actually implies that at large $Q$ the pion transition form
factor does not go to zero. These arguments suggest that pQCD might in
fact be computing just the subtracted form factor. Clearly, the issue
cannot be settled unless one could go smoothly over all the available
energy range.  Of course, quark models are not QCD itself, and it is
quite possible that our considerations do not apply directly to the
QCD analyses. However, the present paper unveils a warning that from
purely field-theoretic reasons an asymptotically non-vanishing pion
transition form factor cannot be rejected.

With this finding in mind we analyze in detail the predictions of IVMD
and the Regge models for $F_{\pi^0 \gamma \gamma^\ast}$. Regge models
with (infinitely many) tree-level meson and glueball exchanges are a
realization of the large-$N_c$ limit \cite{'tHooft:1973jz} which,
unlike the quark models, incorporates confinement. In particular, these
models where insightful in analyzing condensate
issues~\cite{Golterman:2001nk,Beane:2001uj,Beane:2001em,Simonov:2001di,Golterman:2002mi,Afonin:2003gp,RuizArriola:2006gq,Arriola:2006sv,Afonin:2006sa}, form 
factors~\cite{RuizArriola:2006ii,RuizArriola:2008sq}, or the nature of scalar
states~\cite{Arriola:2010fj}.
 
On purely phenomenological grounds, we will show that the data for
$F_{\pi^0 \gamma \gamma^\ast}$ can be fitted very well in a simple
vector-meson exchange model with IVMD. In more elaborate variants of
the Regge approach the data may be reproduced (with different choices
of the model parameters) almost equally well with or without the fulfillment of 
the TW bounds.

There has also been an important interest in the pion transition form
factor after the proposal of determining the rare $Z \to \pi^0 \gamma$
decay~\cite{Jacob:1989pw}, which probes the time-like value
$q^2=M_Z^2$.  The experimental bound 
%$\Gamma( Z \to \pi^0 \gamma) < 5
%\times 10^ {-5} \Gamma_{\rm tot} (Z)= 10.25 \times 10^{-5} {\rm GeV}$,
given by the Particle Data Group~\cite{Amsler:2008zzb} implies a large
suppression as compared to the $q^2=0$ anomaly point, but not
necessarily as much as predicted by many authors, where $\Gamma( Z \to
\pi^0 \gamma) \sim 10^{-11} {\rm GeV}$ (see,
e.g.,~\cite{Bando:1993ct}).

The paper is organized as follows. As a preparatory material, in Sec.~\ref{sec:kin} we provide
some basic kinematics, in Sec.~\ref{sec:TW} we review the TW bounds, and in Sec.~\ref{sec:Z} recall the bound from the 
$Z\to \pi^0 \gamma$ decay. In Sec.~\ref{sec:QM} we use several variants
of the quark model to inquire on the validity of the TW
bounds. As we will show, these models are not capable of describing the
data with realistic model parameters, but pose a warning on the second TW
bound. We use this insight in Sec.~\ref{sec:VMD} to propose a more
realistic Vector Meson Dominance description of the data. The analysis
is complemented in Sec.~\ref{sec:regge} with the inclusion of the
large $N_c$ motivated infinite tower of radial  Regge like
excitations of the $\rho-$meson. The important role of photon
asymmetry is discussed briefly in Sec.~\ref{sec:asym}. Finally, in
Sec.~\ref{sec:concl} we list our main conclusions.

\section{The kinematics \label{sec:kin}}

\begin{figure}[tbc]
\includegraphics[width=0.35\textwidth]{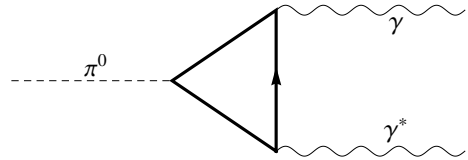} 
\caption{The triangle diagram used to evaluate the pion transition form factor in quark models, with one real and one virtual photon 
(the crossed diagram not shown).
\label{fig:tri}}
\end{figure}

With the outgoing momenta and polarizations of the photons denoted as
$q_1$,$e_1$ and $q_2$, $e_2$ (see Fig.~\ref{fig:tri}), one finds the amplitude
\begin{eqnarray}
\Gamma^{\mu \nu}_{\pi^0  \gamma^\ast \gamma^\ast } (q_1,q_2)  &=& 
\epsilon_{\mu\nu \alpha \beta}e_1^\mu e_2^\nu q_1^\alpha q_2^\beta 
F_{\pi \gamma^\ast \gamma^\ast}(Q^2,A),
\end{eqnarray} 
where the pion transition form factor $F_{\pi \gamma^\ast
\gamma^\ast}$ depends on the total virtuality, $Q^2$, and the photon momentum
asymmetry parameter, $A$,
\begin{eqnarray}
Q^2 = -(q_1^2 + q_2^2 ), \; A = \frac{q_1^2 - q_2^2}{q_1^2+ q_2^2}, \; -1 \le A  \le 1.  
\end{eqnarray} 
Equivalently, $q_1^2 = -\frac{(1+A)}2 Q^2$, $q_2^2=-\frac{(1-A)}2
Q^2$.  For large virtualities, assuming factorization, one finds the standard twist
decomposition of the pion transition form factor \cite{Lepage:1980fj},
\begin{eqnarray}
F_{\pi^0 \gamma^\ast \gamma^\ast} (Q^2, A ) =  J^{(2)} (A)
\frac1{Q^2}  + \dots , \label{twist}
\end{eqnarray} 
with 
\begin{eqnarray} 
\!\!J^{(2)}(A) &=& \frac{4 f}{N_c } \int_0^1 \!\!\! dx
 \frac{\varphi_\pi^{(2)}(x)}{1-(2x-1)^2 A^2}, \label{J2}
\end{eqnarray}
involving the pion parton distribution amplitude PDA, $\varphi(x)$. 

Taking the asymptotic form $\varphi(x)=6x(1-x)$ in Eq.~(\ref{J2})
gives the Brodsky-Lepage asymptotic result~\cite{Lepage:1980fj}
\begin{eqnarray}
 J^{(2)}(A=1)=\frac{6}{N_c f}. \label{BL:as}
\end{eqnarray}

An analysis of the lowest
Gegenbauer moments $a_2$ and $a_4$ of $\varphi(x)$ has been carried out
in Ref.~\cite{Schmedding:1999ap}
and~\cite{Bakulev:2001pa,Bakulev:2002uc,Bakulev:2003cs,Bakulev:2005cp}. Higher
twists have been analyzed in the framework of the light-cone sum
rules~\cite{Agaev:2005rc}.  A direct measurement of the DA has been
presented by the E791 collaboration~\cite{Aitala:2000hb}.  For a
concise review on all these developments see, {\em e.g.},
Ref.~\cite{Bakulev:2004mc} and references therein.

Although theoretically one may take the maximally asymmetric case of
$A=1$, in the experiment this situation is not possible to achieve.
For instance, in the BaBar kinematic setup $-q_1^2 <0.6~{\rm GeV}^2$
and $-q_2^2 >3~{\rm GeV}^2$, suggesting $A \sim 0.9-0.97$. As advanced
in Ref.~\cite{Broniowski:2009ft}, we note that the model results
presented in the following sections become quite sensitive to the
precise value of $A$ at large $Q^2$, hence in precision analyses the
effects of kinematic cuts should be incorporated.

At this point, before undertaking further elaborations, an important
qualification is in order. Eq.~(\ref{J2}) provides a gauge-invariant
high-energy-motivated definition of the pion PDA, which formally determines
a non-perturbative matrix element. A purely low-energy definition of
the PDA involves link operators between the quark fields at different
space-time points and depends on the renormalization scale.  A
non-trivial issue has to do with the equivalence of both definitions
and the scale at which this identification makes sense. We refer to
previous
works~\cite{RuizArriola:2002bp,RuizArriola:2002wr,RuizArriola:2003bs,RuizArriola:2006ii}
for a detailed discussion on this delicate and non-trivial issue. An
outstanding outcome was that in several models 
\begin{eqnarray}
\varphi(x)=1 \label{phi:one}
\end{eqnarray}
at the model low-energy scale; the end-point behavior would be bended after
the QCD evolution to higher scales (see Eq.~(5.26) of Ref.~\cite{Broniowski:2007si} for the detailed 
asymptotic form of the PDA near the end-points after evolution).  
While the identification of the PDA
through Eq.~(\ref{J2}) indeed requires taking asymptotically large
momenta, it does not necessarily follow that the asymptotic expression
itself holds at the so far measured momenta.

\section{Terazawa-West bounds \label{sec:TW}}

Terazawa~\cite{Terazawa:1973hk} and West~\cite{West:1973gd} (for a
review see, e.g.,~\cite{Terazawa:1973tb}) derived unitarity bounds for
the pion transition form factor. The derivation considers the photon
propagator, the charge form factor of the pion, and the transition form
factor. The Schwarz inequality is used to show that
\begin{eqnarray}
{\rm Im} F_{\pi^0 \gamma \gamma^\ast}(q^2) = {\cal O}(1/\sqrt{q^2}) \label{TW:I}
\end{eqnarray}
for time-like momenta, $q^2>4 m_\pi^2$, which we term {\em the first
TW bound}.  Under a further important assumption that there are no polynomial
contributions to the real part of the form factor, one may also
conclude that $\left | F_{\pi^0 \gamma \gamma^\ast}(q^2) \right | =
{\cal O}(1/\sqrt{q^2})$. Then the dispersion relation (see Sec.~14 of
Ref.~\cite{Terazawa:1973tb}) yields that
\begin{eqnarray}
\left | F_{\pi^0 \gamma \gamma^\ast}(Q^2) \right | = {\cal O}(1/Q) \label{TW:II}
\end{eqnarray}
for large space-like momenta $Q$, which we call {\em the second TW bound}. 

The assumption of the absence of the polynomial terms (in particular,
a constant) from the real part of the pion transition form factor is
equivalent to validity of the unsubtracted dispersion relation.  
As mentioned in the Introduction, this
crucial assumption is clearly spelled out as such in the
original papers; in the following Sections of the paper we will
investigate situations where the subtraction constant is present in
the pion transition form factor.

The constant in the bound (\ref{TW:II}) may be given~\cite{Terazawa:1973tb} in terms of the photon spectral density and the
pion quark structure function, namely
\begin{eqnarray}
\left | F_{\pi^0 \gamma \gamma^\ast}(Q^2) \right | < \frac{2 \sqrt{\Pi(\infty)}}{Q} 
\int_0^1 dx \sqrt{\frac{F_1(x,Q^2)}{x(1-x)}}, \label{TW:IIexa}
 \end{eqnarray}
where 
\begin{eqnarray}
\Pi(s) =\frac{s}{16\pi^3 \alpha_{\rm QED}^2} \sigma_{e^+ e^- \to {\rm hadrons}}(s), \label{eq:PI}
\end{eqnarray}
with the asymptotic value 
\begin{eqnarray}
\Pi(\infty) = \frac{1}{12\pi^2} \sum_i e_i^2, \label{PI:infty}
\end{eqnarray}
$e_i$ denoting the charge of the quark of flavor $i$ in the natural units. 
The pion structure function is
\begin{eqnarray}
F_1(x,Q^2)=\frac{1}{2} \sum_i e_i^2 [q_i(x,Q^2) + \bar q_i(x,Q^2)], \label{eq:F1}
\end{eqnarray}
with $q_i(x,Q^2)$ and $\bar q_i(x,Q^2)$ denoting the quark and
anti-quark distribution functions at momentum $Q$.

With the help of the SMRS~\cite{Sutton:1991ay} and
the GRV~\cite{Gluck:1999xe} parameterizations of the pion PDF's we
obtain, via integrating from $x=10^{-5}$ to $1$, a numerical estimate
for the coefficient in the bound (\ref{TW:IIexa}) which actually
exhibits a rather mild dependence on $Q^2$.  With $Q^2$ in the range
$10-40~{\rm GeV}^2$ we find for the LO and NLO evolved PDF's the bounds
\begin{eqnarray}
|F_{\pi^0 \gamma \gamma^\ast}(Q^2)| <&& \frac{0.85(1)}{Q} \quad {\rm
 (LO)} \, , \nonumber \\ < && \frac{0.75(1)}{Q} \quad {\rm (NLO)} \, ,
\end{eqnarray}
with the small uncertainty reflecting the considered $Q$ interval.  Therefore the
bound is completely ``inefficient'' in the considered range of
momenta, as it is an order of magnitude above the BaBar data. For that
reason it is of formal rather than practical interest.~\footnote{We are
  taking $Q^2$ large enough as to neglect the higher twist
  corrections. Actually the bound never crosses the pQCD limit $2
  f_\pi/Q^2$ for the lowest momenta considered in Ref.~\cite{Gluck:1999xe}, namely
  $Q_0^2 \sim 0.26~{\rm GeV}^2$ at LO and $Q_0^2 \sim 0.40~{\rm GeV}^2$ at
  NLO.}

\section{Bound from the $Z \to \pi^0 \gamma$ decay \label{sec:Z}}

Another bound for the pion transition form factor comes from the rare $Z \to
\pi^0 \gamma$ decays. In that process, proceeding via a quark loop,
only the vector coupling of the $Z$ boson to the quarks
contributes~\cite{Jacob:1989pw}, hence
\begin{eqnarray}
\frac{F_{Z \to \pi^0 \gamma} (q^2)}{F_{Z \to \pi^0 \gamma} (0)}=
\frac{F_{\pi^0 \gamma^\ast \gamma} (q^2)}{F_{\pi_0 \gamma^\ast \gamma} (0)}.
\end{eqnarray}
The experimental limit $\Gamma( Z \to \pi^0 \gamma) < 5
\times 10^ {-5} \Gamma_{\rm tot} (Z)= 10.25 \times 10^{-5} {\rm GeV}$,
provided by the Particle Data Group~\cite{Amsler:2008zzb}, implies  
\begin{eqnarray}
|F_{Z \to \pi^0 \gamma} (M_Z^2)/F_{Z \to \pi^0 \gamma} (0)| < 0.17. \label{bound:Z}
\end{eqnarray}
This experimental bound is not as stringent as predicted by many
authors, where $\Gamma( Z \to \pi^0 \gamma) \sim  10^{-11} {\rm
  GeV}$ (see, e.g.,~\cite{Bando:1993ct}).  The bound can be used in
models which predict the form factor in the time-like region, {\em
  e.g.} the Regge models considered in the following Sections.

\section{Quark models \label{sec:QM}}

The role of the constituent vs current quarks in the $\pi^0$ and $\omega$
transition form factors in triangle diagrams was recognized in
Ref.~\cite{Bando:1993cu}. The diagram of Fig.~\ref{fig:tri} is superficially
linearly divergent, but the requirement of the gauge invariance guarantees
convergence. 
In the constituent quark model (CQM) a direct computation of the triangle 
yields, for the general kinematic case, 
\begin{eqnarray}
F_{\pi^0 \gamma^\ast \gamma^\ast} (Q^2,A) = \frac{1}{4 \pi^2 f_\pi} G(Q^2,A),
\end{eqnarray}
where the loop function (we work in the strict chiral limit of $m_\pi=0$), 
\begin{eqnarray}
G(Q^2,A) &=& \frac{2 M^2}{Q^2} \int_0^1 \frac{dx }{x} \times \\ &&
 \log\left[\frac{2M^2 + (1+A) x (1-x) Q^2}{2M^2+ (1-A) x (1-x) Q^2} \right], \nonumber
\label{eq:G}
\end{eqnarray}
is normalized to unity at the origin, $G(0,A)=1$. 
For the special case $A=1$ 
\begin{eqnarray}
&& G(Q^2,A=1) \equiv G(Q^2) = \nonumber \\
&& \frac{2 M^2}{Q^2} \int_0^1 \frac{dx }{x} \log\left[1+x(1-x)
  \frac{Q^2}{M^2} \right]. \nonumber \\
\label{eq:G1}
\end{eqnarray}

We note that the evaluation is covariant, thus different in philosophy
from the conventional light-cone analysis. In the latter case the soft
part of the diagram containing the PDA is factorized, while the hard
part of the diagram is evaluated according to the light-cone Feynman
rules. On the other hand, in CQM one evaluates the diagram of
Fig.~\ref{fig:tri} with the instant-form Feynman rules, corresponding
to local or non-local variants of the model.  That way one is
insensitive to the issues of factorization.  However, it is not clear
how credible this approach is at high virtualities, in particular in
local models, where the virtual quark propagator carries, even
asymptotically, a large constituent mass. In the study of this Section
we are, however, primarily interested in formal aspects, namely, the TW
bounds in a field-theoretic model, thus we shall not be concerned
with the issue whether or not a low-energy quark model can be
realistically used to obtain the asymptotic transition form factor according to
Fig.~\ref{fig:tri}.

Chiral quark models are particular realizations of the large-$N_c$
limit. A variant of the CQM, the well-known Georgi-Manohar
model~\cite{Manohar:1983md}, allows the quarks to carry an axial
charge different from unity, $g_A^Q \neq 1 $.  The relevant part of
the Lagrangian of the model is
\begin{eqnarray}
&& \hspace{-3mm}  L=\bar q \left ( i \slashchar{\partial} + g_A^Q \slashchar{A} \gamma_5  - M   \right ) q 
+ \frac{f^2}{4} {\rm Tr} \left ( \partial_\mu U^\dagger \partial^\mu U \right ) +{\rm WZW}, \nonumber \\
&& \hspace{-3mm} A_\mu=\frac{i}{2}(u^\dagger \partial_\mu u-u \partial_\mu u^\dagger), \;\; u=e^{i \vec{\pi}\cdot\vec{\tau} /(2f)}, 
\;\; U=u^2, \label{eq:gm}
\end{eqnarray}
where $f=93~{\rm MeV}$ is the pion decay constant and WZW denotes the
Wess-Zumino-Witten term~\cite{Wess:1971yu,Witten:1983tx}.  There has
been some discussion on whether or not $1-g_A^Q = {\cal O}(N_c^0 )$.
This issue was answered in the affirmative way in
Ref.~\cite{Broniowski:1993nb}, where the ${\cal O}(N_c^0)$ departure
from unity is basically due to the $t$-channel exchanges in the
Adler-Weisberger sum rule for the pion-quark scattering. In the NJL
model with vector mesons, the $A_1-\pi$ mixing is an explicit
leading-$N_c$
source~\cite{Klimt:1989pm,Vogl:1989ea,RuizArriola:1991gc} of the
effect. Because of the chiral anomaly, the field representation used
to describe the $g_A^Q \neq 1$ situation is relevant. In
Ref.~\cite{Manohar:1984uq} it was shown that the apparent anomalous
inequivalence of effective theories may be compensated by including
extra terms in the effective action. More specifically, the original
GM model {\it does not} contain the anomalous piece, such that the
Wess-Zumino-Witten (WZW) action should be added {\it a posteriori}, as
in Eq.~(\ref{eq:gm}). An example on how various vertex functions are
modified by naively including $g_A^Q \neq 1 $ and violating the
anomaly in chiral quark models is presented in
Ref.~\cite{RuizArriola:1993sp}. The subtraction procedure described
above to restore the anomaly and based on \cite{Manohar:1984uq} was
addressed in Ref.~\cite{Bijnens:1993cy} (see also
\cite{RuizArriola:1995ea}).

A direct calculation of the pion transition form factor 
in the GM model supplied with the WZW action yields the result
\begin{eqnarray}
F_{\pi^0 \gamma \gamma^*} (Q^2) = \frac{1}{4 \pi^2 f_\pi} + \frac{g_A^Q}{4
  \pi^2 f_\pi} \left[ G(Q^2)-1 \right],
\label{eq:f-ga}
\end{eqnarray}
with $G(Q^2)$ given by Eq.~(\ref{eq:G}). This expression satisfies the
anomaly and the dispersion relation but {\it does not} vanish at
infinity, since
\begin{eqnarray}
F_{\pi^0 \gamma \gamma^*} (Q^2) = \frac{1-g_A^Q}{4 \pi^2 f_\pi}  + 
\frac{g_A^Q M^2}{4 \pi^2 f_\pi} \frac{\left[\log (Q^2/M^2)\right]^2}{Q^2} + \dots \nonumber \\
\end{eqnarray}
fulfills the first TW bound (for the imaginary part) but {\it does not} fulfill the second
bound (for the real part). 

Similarly, within SQM~\cite{RuizArriola:2003bs} a direct
implementation of $g_A^Q$ is equally possible and one gets 
%in the rotated basis 
the result of Eq.~(\ref{eq:f-ga}) with 
\begin{eqnarray}
G(Q^2)= \frac{1}{3}\left[ \frac{2 m_\rho^2}{m_\rho^2+Q^2}+\frac{m_\rho^2}{Q^2} \log
\left ( \frac{m_\rho^2+Q^2}{m_\rho} \right ) \right].
\label{eq:fpigg}
\end{eqnarray}
With $g_A^Q = 1 $ this model fulfills qualitatively the result of
Radyushkin with a similar mass scale~\cite{Radyushkin:2009zg}, since
\begin{eqnarray}
F_{\pi^0 \gamma \gamma^*} (Q^2) = \frac{1-g_A^Q}{4 \pi^2 f_\pi} +
\frac{g_A^Q m_\rho^2}{12 \pi^2 f_\pi} \frac{\left[\log
    (Q^2/m_\rho^2)\right]}{Q^2} + \dots \nonumber \\
\end{eqnarray}

In the {\it local} quark models we have the general relation 
%(extended for the $g_A^Q \neq 1$ case)
\begin{eqnarray}
F_{\pi^0 \gamma \gamma^*} (t) &=& \frac{1-g_A^Q}{4 \pi^2 f_\pi} +
\frac{g_A^Q}{12 \pi^2 f } \left[2 F_{\pi^+}^{\rm em}(t) \right . \nonumber \\ 
&& \left . + \int_0^t ds F_{\pi^+}^{\rm em}(s) \right],
\label{eq:fv-fpig}
\end{eqnarray}
which correlates the charge and transition form factors. This relation
shows that in these models even in the case $g_A^Q-1$ one cannot have complete VMD simultaneously for the two 
form factors. 

Unfortunately, attempts to fit the experimental data with the
quark-model formulas, with $g_A$ equal or different from unity, are
not numerically successful in the whole momentum range, 
unless one uses unrealistic model parameters.  The
purpose of the above calculations was different.  Our examples show in
an explicit manner that the second TW may be violated in models
consistent with all field-theoretic constraints, such as covariance,
gauge invariance, chiral symmetry, and anomaly matching. Importantly,
they provide instances where the form factor does not vanish
asymptotically. The calculation also illustrates that the $Q^2=0$
value and the finite $Q^2$ values are independent.  The possibility of
disobeying the second TW bound will be explored in the following
Sections.

\section{Vector meson dominance models \label{sec:VMD}}

We now come to the core of our paper and consider several
implementations of IVMD, both in a simple approach with a single
vector meson, as well as in more sophisticated Regge models with
infinitely-many radially-excited states.  It will turn out that the pion transition
form factor can be accurately described with this phenomenological
method. The coupling of photons to vector mesons has a long history
and Ref.~\cite{O'Connell:1995wf} comprehensively reviews the interplay
between universality, the vector-meson dominance and the low-energy
theorems. We will highlight first the issue of IVMD for the charge
form factor, such that our points are later on more easily made for the transition
form factor.

\subsection{Charge form factor}

The charge form factor is the famous case where VMD can be implemented:
\begin{eqnarray}
F_{\pi^+}^{\rm em}(t)= \frac{M_V^2}{M_V^2-t} \to
-\frac{M_V^2}{t} + \dots \label{s:vmd}
\end{eqnarray}
with $t=-Q^2$.
When the unregularized quark-loop mean squared
radius~\cite{Tarrach:1979ta} is matched to (\ref{s:vmd}), one gets the
relation
\begin{eqnarray}
M_V^2 = 24 \pi^2 f_\pi^2 /N_c.  
\label{eq:vmd-qm-Mv}
\end{eqnarray}
There is no way to match the successful form factor of
Eq.~(\ref{s:vmd}) to the pQCD result.\footnote{If we nevertheless match
  Eq.~(\ref{s:vmd}) to pQCD, the only possible solution is
  $\alpha_s(Q) = \pi/2$, which yields a too small scale $Q \sim 300~
  {\rm MeV}$. This matching to pQCD may seem weird but need not
  necessarily be conceptually wrong. Note that in the string model
  calculation of the $q \bar q $ potential one gets $V_{\bar q q }
  (r)= - \pi/12 r $, whereas from pQCD $V_{\bar q q } (r)= - 4
  \alpha_S/3 r $. This yields $\alpha = \pi/16$, which means a safely
  high scale of $\mu = 2~{\rm GeV}$ for $\Lambda_{QCD}=240~{\rm
    MeV}$. The string model describes accurately the lattice data in the
  short distance region.  In the case of the charge form factor with
  VMD, the required scale is much smaller.}

If one uses a once-subtracted dispersion relation and imposes the current
conservation, $F_{\pi^+}^{\rm em} (0) =1$, one gets 
\begin{eqnarray}
F_{\pi^+}^{\rm em}(t) - 1 = \frac1\pi \int_{t_0}^\infty \frac{t}{t'}\frac{{\rm Im} F_{\pi^+}^{\rm em} (t')}{t'-t-i\epsilon} dt' ,
\label{eq:dr-em}
\end{eqnarray} 
which yields 
\begin{eqnarray}
F_{\pi^+}^{\rm em}(-\infty) - 1 = -\frac1\pi \int_{t_0}^\infty \frac{{\rm Im} F_{\pi^+}^{\rm em} (t')}{t'} dt' .
\end{eqnarray} 
Saturation with a single resonance gives the result
\begin{eqnarray}
F_{\pi^+}^{\rm em}(t)= 1+ \frac{a}{2} \frac{t}{M_V^2-t} \to 1-\frac{a}{2} + 
\frac{a}{2}\frac{M_V^2}{-t}  + \dots 
\label{eq:fpi-em}
\end{eqnarray}
When the coupling $a/2= f_{\rho \gamma} g_{\rho \pi \pi} /m_\rho^2
\equiv g_{\rho \pi \pi}/g_{\rho e^+ e^-} $ becomes unity, the form
factor vanishes at infinity. Sakurai's universality indeed requires $g_{\rho
  \pi \pi}= g_{\rho e^+ e^-} $. In general, however, we may have $g_{\rho \pi \pi}
\neq g_{\rho e^+ e^-} $. 
%The available data suggest an increasing trend. 
Note that positivity requires $a< 2$, such that $F_{\pi^+}^{\rm
  em}(-Q^2) < 1$.  A fit to the experimental data
\cite{Bebek:1977pe,Volmer:2000ek,Horn:2006tm,Tadevosyan:2007yd} yields
$a/2=0.99(2)$ and $M_V =679(36)~{\rm MeV}$.  Thus the data for the
pion charge form factor are consistent with the complete VMD, but
certainly do not preclude IVMD, with $a/2$ departing slightly from
unity.

Caldi and Pagels~\cite{Caldi:1976gz} obtained a similar expression as
Eq.~(\ref{eq:fpi-em}) for the pion form factor from a direct photon
contribution and a momentum dependent $\rho-\gamma$ vertex (see also
Ref.~\cite{O'Connell:1995xx}). The discussed properties are nicely
displayed by the hidden-symmetry approach by Bando {\it et
  al.}~\cite{Bando:1984ej}, where there is a contact piece and the
vector meson term, which dominates completely when the KSFR relation
is fulfilled. Actually, in Ref.~\cite{Benayoun:1998ss} the equivalence
of this approach to more conventional ones is established.  The
interplay between VMD and universality was analyzed by
Schechter~\cite{Schechter:1986vs} (see also
Ref.~~\cite{Zerwekh:2006tg}). Symmetry breaking effects have been
analyzed in Ref.~\cite{Harada:1995sj}. More attempts including
predictions for meson decays can be found in~\cite{Klingl:1996by}.

Of course, the non vanishing of the low-energy representation (the 
one-resonance saturation) need not be taken as a fundamental problem. The
only feature we see is that this non-vanishing represents more
accurately the unknown high-energy data. If we had infinitely
many states, we could fit that data and by separating explicitly the
contribution from the lowest $\rho$-state. We see that the effect
of all other states does behave as the $a$ constant, which is slowly
dependent on $Q$ in a {\it wide} energy range.	

Ideally, one should take the $e^+ e^- \to \pi^+ \pi^-$ data over all possible
momenta. However, obviously the experimental data are  available only up to a
certain maximum value, $s_{\rm max}= 4 \Lambda^2 $. Thus, even if the form
factor vanishes at infinity, we have
\begin{eqnarray}
F_{\pi^+}^{\rm em}(t) - 1 &=& \frac1\pi \int_{4 m_\pi^2}^{4
  \Lambda^2}\ \frac{t}{s}\frac{{\rm Im} F_{\pi^+}^{\rm em} (s)}{s-t}
ds \nonumber \\ && + \frac1\pi \int_{4 \Lambda^2}^\infty\ \frac{t}{s}\frac{{\rm Im}
  F_{\pi^+}^{\rm em} (s)}{s-t} ds, 
\label{eq:dr-em2}
\end{eqnarray} 
where the last term is weakly momentum-dependent and hence resembles a
constant behavior assumed in IVMD, as discussed above.

\subsection{Transition form factor \label{sec:IVMD}}

For the $\pi^0 \gamma \gamma^* $ form factor the complete VMD with just one state implies
\begin{eqnarray}
F_{\pi^0 \gamma \gamma^*}(t)= \frac{1}{4 \pi^2 f} \frac{M_V^2}{M_V^2-t} \to
%-\frac{6f_\pi}{N_c t} 
-\frac{M_V^2}{4\pi^2 f t}
+ \dots \label{s:tf}
\end{eqnarray}
Note that while the anomaly value does not depend on $N_c$ explicitly,
the high momentum behavior does. When matching Eq.~(\ref{s:tf}) to
the pQCD result of Eq.~(\ref{BL:as}) is done, one gets independently the relation (\ref{eq:vmd-qm-Mv}),
\begin{eqnarray}
M_V^2 = 24 \pi^2 f_\pi^2 /N_c . 
\end{eqnarray}
%which for $N_C=3$ recovers the QM-VMD relation which can also be
%obtained by a similar matching of the electromagnetic form factor in the quark model. 
Despite this appealing property, the parametrization (\ref{s:tf})
fails to describe the experimental data in the high momentum region.

If we incorporate the possibility that the form factor need not vanish
at infinity, we may write a once-subtracted dispersion
relation~\cite{Truong:2001qi},
\begin{eqnarray}
\hspace{-2mm} F_{\pi^0 \gamma \gamma^*}(t) - F_{\pi^0 \gamma \gamma^*}(0) = 
\frac1\pi \int_{t_0}^\infty \frac{t}{t'}\frac{{\rm Im} F_{\pi^0 \gamma \gamma^*}(t')}{t'-t-i\epsilon} dt'  .
\label{eq:dr-anom}
\end{eqnarray} 
The influence of the well-known time-like region does not determine
unambiguously when the onset of the pQCD takes place.  Actually, the
single VMD model shows that even in the space-like region with momenta as low as
$Q^2 \sim m_\rho^2$ the effects of the chiral logs and final state interactions are
meager. Taking the limit $t \to -\infty$ we get 
\begin{eqnarray}
F_{\pi^0 \gamma \gamma^*}(-\infty) - F_{\pi^0 \gamma \gamma^*}(0) = 
-\frac1\pi \int_{t_0}^\infty \frac{{\rm Im} F_{\pi^0 \gamma \gamma^*}(t')}{t'} dt'  . \nonumber \\
\label{eq:dr-2}
\end{eqnarray} 
This shows that $F_{\pi^0 \gamma \gamma^*}(-\infty) < F_{\pi^0 \gamma
  \gamma^*}(0)$ if ${\rm Im} F_{\pi^0 \gamma \gamma^*}(t) > 0$.
When we saturate the absorptive part with just one resonance, we get 
\begin{eqnarray}
F_{\pi^0 \gamma \gamma^*} (-Q^2) = \frac{1}{4 \pi^2 f_\pi} \left[ 1 -
  c \frac{Q^2}{M_V^2+Q^2} \right] , \label{form:ivmd}
\end{eqnarray}
with $M_V$ denoting the vector-meson mass.
The coefficient $c$ is related to the $\rho \to \pi \gamma $ decay,
\begin{eqnarray}
c=c_{\rho \pi \gamma}= \frac{-2 e f_\rho}{m_\rho} \frac{{\cal A}(\rho^0
  \to \pi^0 \gamma)}{{\cal A}(\pi^0 \to \gamma \gamma)} = 1.022 \pm 0.051, \label{cest}
\end{eqnarray}
where the value of the estimate is obtained from the decay width $\Gamma (\rho^+ \to \pi^+ \gamma) = 68 \pm 7 {\rm KeV}$. 

\begin{figure}[tbc]
\includegraphics[width=0.45\textwidth]{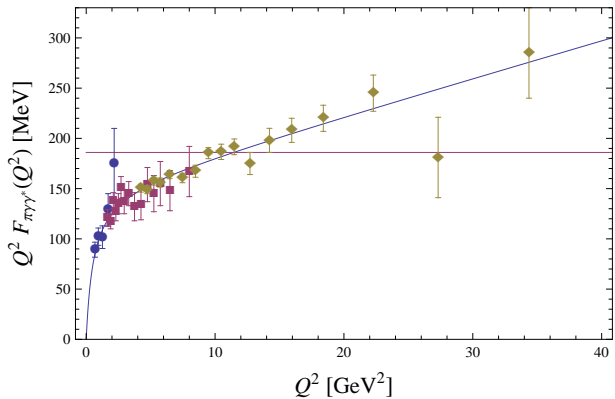} 
\caption{The pion transition form factor in the space-like region $Q^2=-t > 0$. The dots, squares, and diamonds correspond to the 
CELLO~\cite{Behrend:1990sr}, CLEO~\cite{Gronberg:1997fj}, and BaBar~\cite{Aubert:2009mc} data, correspondingly. 
The line is the incomplete-vector-meson-dominance fit
with formula (\ref{form:ivmd}) and parameters (\ref{ivmd:fit}). 
\label{fig:ftpQCD}}
\end{figure}

\begin{figure}[tbc]
\includegraphics[width=0.4\textwidth]{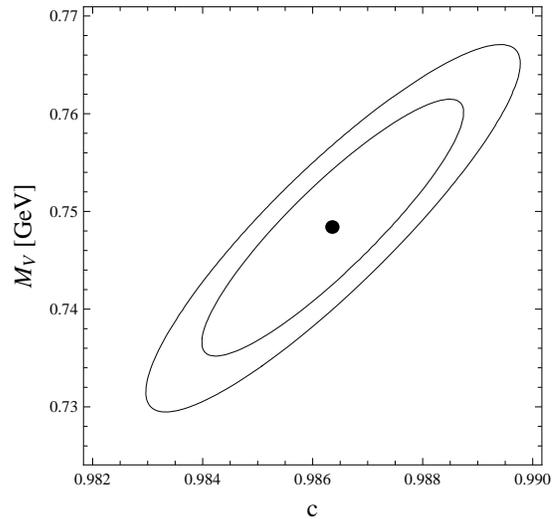} 
\caption{The $\Delta \chi^2 = 2.3$ and $4.6$ contours, corresponding
  to the $68\%$ and $90\%$ confidence levels, in the $M_V-c$ plane
  for the fit with the IVMD ansatz (\ref{form:ivmd}). The central values yield $\chi^2 /{\rm DOF}=0.6$.
\label{fig:ftpQCD-contour}}
\end{figure}

If we fit the parameters in formula (\ref{form:ivmd}) with the CLEO data only, 
%with $A=1$ 
we get
\begin{eqnarray}
c = 0.998(18), \;\;\; M_V=777(44)~{\rm MeV}, \label{fit:cleo}
\end{eqnarray}
with $\chi^2/{\rm DOF}=0.54, $ hence $c$ is consistent with unity, in
agreement with previous determinations assuming complete VMD, however
not excluding IVMD. On the other hand, the fit to the combined CELLO,
CLEO, and BaBar data yields the result
\begin{eqnarray}
c=0.986(2), \;\; M_V=748(14){\rm MeV}, \label{ivmd:fit} 
\end{eqnarray}
hence $c$ is different form unity at the level of 7 standard
deviations, while it remains consistent with estimate (\ref{cest}).
As seen from Fig.~\ref{fig:ftpQCD}, the agreement with the
experimental data over all the momentum range is remarkable, with
$\chi^2/{\rm DOF}=0.7$.  The contours of the confidence-levels are
displayed in Fig.~\ref{fig:ftpQCD-contour}.

%Note that the $Q^2$ term is independent of the asymmetry $A$, as is also apparent from Fig.~\ref{fig:pex}. 

The logarithmic slope at the origin reads 
\begin{eqnarray} 
b_\pi &=& -\left[\frac1{F_{\pi^0 \gamma \gamma^\ast} (Q)}
\frac{d}{d Q^2} F_{\pi^0 \gamma \gamma^\ast} (Q)
\right]\Big|_{Q^2=0}.
\end{eqnarray}
Numerically, taking the optimum parameters (\ref{ivmd:fit}), we get the value 
\begin{eqnarray} 
b_\pi = \frac{c}{M_V^2} = 1.76(7)~{\rm GeV}^{-2} \, .   
\end{eqnarray}
Our IVMD model estimate is 
in a very good agreement with the average experimental values quoted in
the PDG~\cite{Amsler:2008zzb}: $b_\pi = (1.76 \pm 0.22 ) {\rm
  GeV}^{-2}$. This value is very close to that originally reported by
the CELLO collaboration~\cite{Behrend:1990sr}, obtained from an
extrapolation from high-$Q^2$ data to low $Q^2$ by means of
generalized vector meson dominance, $b_\pi = (1.4 \pm 1.3 \pm 2.6)
{\rm GeV}^{-2}$ given in \cite{Farzanpay:1992pz}, and $b_\pi = (1.4
\pm 0.8 \pm 1.4) {\rm GeV}^{-2}$ given in \cite{MeijerDrees:1992qb}.

To end this Section, we turn to the rare decay $Z \to \pi^0 \gamma$,
which probes the transition form factor in the physical region. From
Eq.~(\ref{form:ivmd}) with parameters (\ref{ivmd:fit}) we get $|F_{Z
\to \pi^0 \gamma} (M_Z^2)/F_{Z \to \pi^0 \gamma} (0)| = 0.014(2)$, a
comfortable order of magnitude smaller than the upper experimental
bound (\ref{bound:Z}),
but not as small as predicted by several models ($\sim 10^{-4}$, see,
e.g.,~\cite{Bando:1993ct} and references therein).

\section{Regge models \label{sec:regge}}

In the previous Section we have considered the simplest possible
implementation of IVMD, with just one vector-meson state. However, the
large-$N_c$ limit of QCD involves tree-level diagrams with
infinitely many states, including the radial excitations. That way the matching to the QCD correlators
can be accomplished
\cite{RuizArriola:2006gq,Arriola:2006sv,Arriola:2010fj}, as well as the
correct asymptotic behavior of the pion charge form
factor~\cite{RuizArriola:2008sq} may be obtained.  
In this Section we analyze the pion
transition form factors in the framework of Regge models with
infinitely many radially excited vector-meson states.

\begin{figure}[tbc]
\includegraphics[width=0.185\textwidth]{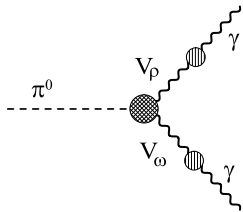} 
\caption{The pion transition form factor in Regge models. 
The labels $V_\rho$ and $V_\omega$ denote the infinite vector meson towers  with 
the $\rho$ and $\omega$ quantum numbers. 
\label{fig:reggev}}
\end{figure}

Based on the success of the Veneziano-Lovelace-Shapiro dual resonance
model (see, e.g., \cite{Veneziano:1974dr,Mandelstam:1974fq} and
references therein) Suura~\cite{Suura:1969eu} and
Frampton~\cite{Frampton:1969ry} proposed analytic models. The general
form reads
\begin{eqnarray}
F_{\pi^0 \gamma^\ast \gamma^\ast} (Q^2,A) &=& 
\sum_{V_\rho,V_\omega} \frac{F_{V_\rho}(q_1^2) F_{V_\omega}(q_2^2) 
G_{\pi V_\rho V_\omega}(q_1^2,q_2^2)}{(q_1^2 - M_{V_\rho}^2)(q_2^2 -M_{V_\omega}^2 )}
\nonumber \\ &+& (q_1 \longleftrightarrow q_2), \label{eq:amp}
\end{eqnarray} 
where $F_{V_\rho}$ and $F_{V_\omega}$ are the current-vector meson
couplings, while $G_{\pi V_\rho V_\omega}$ is the coupling of two vector
mesons to the pion. The situation is depicted in Fig.~\ref{fig:reggev}. At the soft photon point, corresponding to the
neutral pion decay $\pi^0 \to 2 \gamma $, the chiral anomaly matching
condition imposes the normalization
\begin{eqnarray}
F_{\pi^0 \gamma^* \gamma^* } (0,0) &=& \sum_{V_\rho V_\omega} 
\frac{2F_{V_\rho}(0) F_{V_\omega}(0) G_{\pi V_\rho V_\omega}(0,0)}
{M_{V_\rho}^2 M_{V_\omega}^2} \nonumber \\ &=& \frac{1}{4 \pi^2 f}. 
\label{anomc}
\end{eqnarray} 
This consistency constraint, realized in nature, can be always
satisfied in models by an appropriate choice of the couplings.  In
Ref.~\cite{RuizArriola:2006ii} we analyzed the transition form factor
with the help of formula (\ref{eq:amp}) with the coupling constants
$F_V$ taken as constants (as requested asymptotically by the matching
of spectral densities to QCD) and allowing for constant diagonal
couplings in $G_{\pi V_\rho V_\omega}$.  As a result, a constant PDA of Eq.~\ref{phi:one}
was extracted. The corresponding pion transition form factor has the
asymptotic behavior $\sim \log(Q^2/\mu^2) /Q^2$, however with physical
values of the model parameters it overshoots the data in the region
above $Q^2 \sim 1~{\rm GeV}^2$, thus does not properly reproduce the
data.\footnote{As already mentioned, we stress that taking the $Q^2
  \to \infty$ limit is an operational way of extracting the leading
  twist PDA; this is different than describing the current data within
  this limit.}

In the analysis below we use a more sophisticated approach, namely the
factorizable product suggested by Dominguez~\cite{Dominguez:1983aa} in
the dual description of radiative decays, 
\begin{eqnarray}
F_{\pi^0 \gamma \gamma^*}(t)= \frac{1}{4 \pi^2 f_\pi } f_b(t), \label{eq:reg}
\end{eqnarray} 
where 
\begin{eqnarray}
f_b(t) &=&\frac{1}{B ( b -1 , \frac{M_V^2}{a} )} \label{beta} \\ &\times& \sum_{n=0}^\infty
\frac{\Gamma(2-b+n)}{\Gamma(n+1) \Gamma(2-b)} \frac1{a n + M_V^2 -t}. \nonumber
\label{eq:fb}
\end{eqnarray} 
The function $f_b(t)$ depends on three parameters: the lowest-lying
vector-meson mass, $M_V$, the string tension, $\sigma = a /(2 \pi)$, and
the asymptotic fall-off parameter, $b$. The mass formula is then
$M_V(n)^2= a n + M_V^2$.  The function (\ref{eq:fb}) fulfills the
normalization condition
\begin{eqnarray}
f_b(0)=1.
\end{eqnarray}  
For  $x \gg y $  one has  $B(x,y) \sim \Gamma(y ) x^{-y}$, hence in the 
asymptotic region of $ M_V^2 - t \gg (b-1) a$ we find
\begin{eqnarray}
f_b(t) \sim \frac{\Gamma\left( \frac{M_V^2}a +b-1 \right)}{\Gamma\left(
\frac{M_V^2}a\right)} \left( \frac{a}{M_V^2-t} \right)^{b-1}.
\end{eqnarray} 
The TW bounds are satisfied if $b>1.5$.

We remark that in Ref.~\cite{RuizArriola:2008sq} this version of the Regge approach was used 
to describe the charge form factor. We have shown that it can be accurately reproduced up to
$Q^2 \sim 6{\rm GeV}^2$, while the pQCD result greatly undershoots the
experiment. The onset of pQCD occurs at extremely high (``cosmological'')
values of $Q^2$. 

\begin{figure}[tbc]
\includegraphics[width=0.45\textwidth]{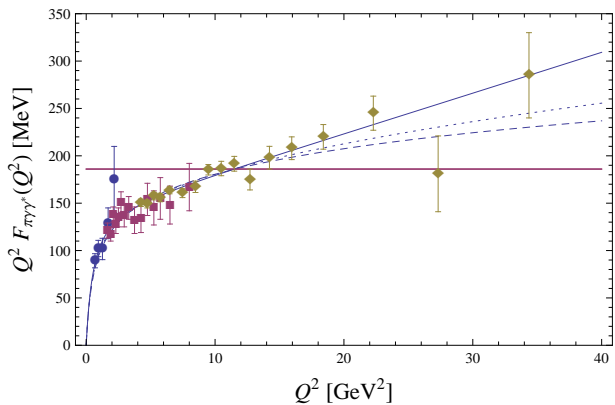} 
\caption{The pion transition form factor in the space-like region $Q^2=-t > 0$. The dots, squares, and diamonds correspond to the 
CELLO~\cite{Behrend:1990sr}, CLEO~\cite{Gronberg:1997fj}, and BaBar~\cite{Aubert:2009mc} data, correspondingly. 
The dashed line is the results of the Regge fit with formula (\ref{eq:reg}) and parameters (\ref{reg:1}). 
The dotted line shows the Regge fit with Eq.(\ref{regmod}) and parameters (\ref{reg:3}). Finally, the solid line 
corresponds to the subtracted Regge model of Eq.~(\ref{regsub}) and parameters (\ref{reg:4}).
\label{fig:regge}}
\end{figure}

We now proceed with the application of the presented Regge model to the pion transition form factor.
Taking $a = 1.3~{\rm GeV}^2$ (which correspond to the string tension $\sigma=(455 ~{\rm MeV})^2$), 
a $\chi^2$  fit of formula (\ref{eq:reg}) to the 
joint CELLO, CLEO, and BaBar data yields 
\begin{eqnarray}
M_V=0.672(25) {\rm ~GeV}, \;\; b=1.81(3),  \label{reg:1}
\end{eqnarray}
with $\chi^2 /{\rm DOF}=1$. The fit is shown in Fig.~\ref{fig:regge}
with the dashed line.  Since $b>1.5$, the fit satisfies the TW
bounds. For the logarithmic slope at the origin we find
$b_\pi=2.1(2)~{\rm GeV}^{-2}$. We also get, with smoothing the
spectral density in the physical region, the ratio $|F_{Z \to \pi^0
  \gamma} (M_Z^2)/F_{Z \to \pi^0 \gamma} (0)| = 0.0014(4)$, which is
comfortably below the experimental bound and an order of magnitude
smaller than the IVMD fit of Sect.~\ref{sec:IVMD}.

\begin{figure}[tbc]
\includegraphics[width=0.45\textwidth]{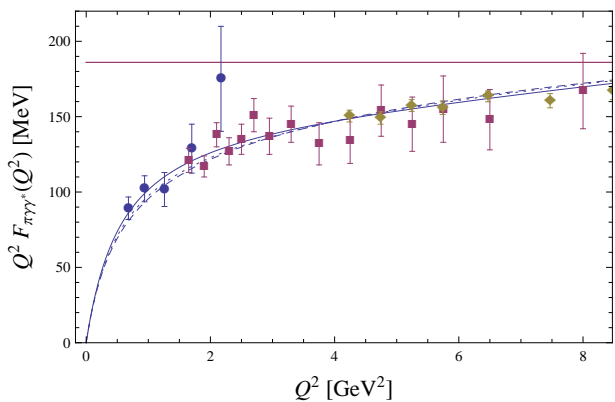} 
\caption{Same as Fig.~\ref{fig:regge} for a smaller range of $Q^2$.
\label{fig:regges}}
\end{figure}

We can further try to improve the agreement with the BaBar data by explicitly separating the first
pole as follows: 
\begin{eqnarray}
F_{\pi^0 \gamma \gamma^*}(t)= \frac{1}{4 \pi^2 f_\pi }\left[ c_\rho
  \frac{m_\rho^2}{m_\rho^2-t} + (1-c_\rho) \frac{\bar f_b(t)}{\bar
    f_b(0)} \right], \label{regmod}
\end{eqnarray}
where $\bar f_b(t) $ is obtained from Eq.~(\ref{eq:fb}) with the $n=0$ term omitted from
the sum. 
We impose the TW bound by setting $b=1.5$. 
Then (with $a=1.3~{\rm GeV}^2$) we find
\begin{eqnarray}
c=0.95(2), \;\; M_V= 709(9)~{\rm MeV}^2, \label{reg:3}
\end{eqnarray}
with $\chi^2 /{\rm DOF}=0.85$, $b_\pi=1.9(1)~{\rm GeV}^{-2}$, and  
$|F_{Z \to \pi^0 \gamma} (M_Z^2)/F_{Z \to \pi^0 \gamma} (0)| = 0.01(1)$.
The corresponding curve is presented in Fig.~\ref{fig:regge} with the dotted line. 

Finally, we take a subtracted Regge model, in analogy to the model of Sec.~\ref{sec:IVMD} of the form 
\begin{eqnarray}
F_{\pi^0 \gamma \gamma^*}(t)= \frac{1}{4 \pi^2 f_\pi }\left[ 1 - c + c f_b(t) \right]. \label{regsub}
\end{eqnarray}
Setting $M_V=770$~MeV and $a=1.3~{\rm GeV}^2$ yields the optimum
values for the remaining parameters
\begin{eqnarray}
c=0.984(4), \;\; b=2.05(3), \label{reg:4}
\end{eqnarray}
with $\chi^2 /{\rm DOF}=0.7$.  Since the value of $b$ is consistent
with $2$, the model gives very similar results to the model with the
single vector-meson state considered in Sec.~\ref{sec:IVMD}. Hence we
find $b_\pi=1.69(1)~{\rm GeV}^{-2}$, and $|F_{Z \to \pi^0 \gamma}
(M_Z^2)/F_{Z \to \pi^0 \gamma} (0)| = 0.021(4)$.  The corresponding
curve is presented in Fig.~\ref{fig:regge} with the solid line.
Within the present model we see that the higher radially excited states of the vector mesons,
$\rho',\rho'', \dots $ and $\omega',\omega'',\dots$, are weakly coupled.

We also zoom the low-$Q^2$ range in Fig.~\ref{fig:regges}. All the
considered fits practically overlap in the displayed range, in
agreement with the observation that IVMD could not be rejected even at
$Q^2 \sim 8 ~ {\rm GeV}^2$. Our analysis shows that higher energy data
do in fact favor IVMD at the level of 7 standard deviations. 

\section{Asymmetry parameter \label{sec:asym}}

\begin{figure}[tbc]
\includegraphics[width=0.45\textwidth]{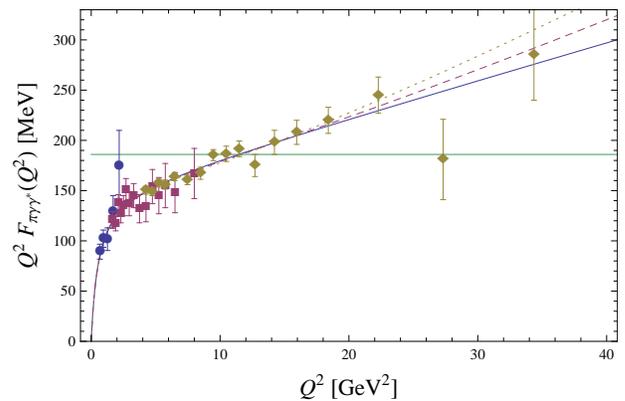} 
\caption{The IMVD fit of Eq.~(\ref{form:ivmdA}) with $A=1$, $0.975$,
  and $0.95$, denoted with solid, dashed, and dotted lines,
  respectively.
\label{fig:ftpQCDA}}
\end{figure}

As mention in the Introduction, in the BaBar kinematic setup $-q_1^2
<0.6~{\rm GeV}^2$ and $-q_2^2 >3~{\rm GeV}^2$, suggesting $A \sim
0.9-0.97$, hence $A$ is not strictly $1$. This departure has
significance for the fits and the obtained parameters, which should
not be forgotten in precision analyses.  We take as an example the
IMVD model, which now becomes
\begin{eqnarray}
&& F_{\pi^0 \gamma \gamma^*} (-Q^2) = \frac{1}{4 \pi^2 f_\pi} \times \\ && \left[ 1 -
  c \left(1- \frac{4M_V^4}{4M_V^4+4M_V^2 Q^2+(1-A^2)Q^4} \right) \right] . \nonumber \label{form:ivmdA}
\end{eqnarray}
Fitting with $A=1$, $0.975$, and $0.95$ yields, respectively,
$c=0.986$, $0.978$, $0.974$, and $M_V=748$, $754$, $768~{\rm MeV}$. We
note significantly different values of the optimum parameters, with
$M_V$ closest to the physical value for $A=0.95$.  The corresponding
curves are displayed in Fig.~\ref{fig:ftpQCDA}.

\section{Conclusions \label{sec:concl}}

These are our main findings:

\begin{itemize}

\item The Terazawa-West bound, asserted for the real part of the pion
  transition form factor, need not be fulfilled in field-theoretic
  approaches. We provide an explicit counter-example with the
  low-energy Georgi-Manohar model, where the chiral anomaly is
  fulfilled but the transition form factor does not vanish at $t \to
  -\infty$.  Provided this feature holds in QCD, it opens a
  possibility of explaining the BaBar experimental data with models
  incorporating the incomplete vector-meson dominance.

\item Moreover, the coefficient in the Terazawa-West bound, estimated
  with the help of a phenomenological parametrization of the pion
  parton distribution functions, is large, extending an order of
  magnitude above the BaBar data. Thus, even if it holds in the
  absence of polynomial contributions to the real part, it is
  completely ineffective for the momentum range of interest.

\item Already the simplest model with the incomplete vector-meson
  dominance, incorporating a single vector-meson state, is capable of
  reproducing the data in the whole available experimental range, $0 <
  Q^2 < 35~{\rm GeV}^2$.

\item Within the Regge approach, where infinitely many radially excited states are
  included, the data can be fitted both ways: satisfying or violating
  the TW bound. The agreement with the experiment is satisfactory,
  both near $Q^2-0$, where the anomaly value and the slope of the form
  factor are reproduced, as well as in the intermediate CLEO range and
  the high-$Q^2$ BaBar range.

\item An additional constraint on the models in the time-like region
  of momenta follows from the rare $Z \to \pi^0 \gamma$ decay.  We use
  this bound in our considerations. For the considered models it is comfortably
  satisfied.

\item Finally, we note that the numerical fits are quite sensitive to the photon momentum asymmetry
  parameter, $A$, which leads to sensitivity in the physical
  parameters, such as the vector meson mass, and sensitivity of the
  transition form factor in the asymptotic range. Since $A$ is not strictly $1$, the effect of the 
  kinematic cuts should be considered in precision analyses.

\end{itemize}

%\bibliography{ivmd-babar,ivmd-babar_0}

\end{document}